\documentclass[preprint,preprintnumbers,amsmath,amssymb,11pt]{revtex4}
\usepackage{amsmath,epsfig,graphicx,amssymb}

\begin{document}
\title{Skyrme-Random-Phase-Approximation description of spin-flip
and orbital giant resonances}

\author{V.O. Nesterenko$^{1}$, J. Kvasil$^{2}$, P. Vesely$^{2,3}$ , W. Kleinig$^{1,4}$,
and P.-G. Reinhard$^{5}$}
\affiliation{$^{1}$\it BLTP, Joint Institute for
Nuclear Research, 141980, Dubna,
Moscow region, Russia\\
nester@theor.jinr.ru; kleinig@theor.jinr.ru}
\affiliation{$^{2}$\it
Institute of Particle and Nuclear Physics, Charles University,
CZ-18000 Praha, Czech Republic\\
kvasil@ipnp.troja.mff.cuni.cz; vesely@ipnp.troja.mff.cuni.cz}
\affiliation{$^{3}$\it
Department of Physics, P.O. Box 35 (YFL),
40014 University of Jyvaskyla,  Jyvaskyla, Finland}
\affiliation{$^{4}$\it
Technische Universit\"at Dresden, Inst. f\"ur
Analysis, D-01062, Dresden, Germany.}
\affiliation{$^{5}$\it
Institut f\"ur Theoretische Physik II, Universit\"at Erlangen, D-91058,
Erlangen, Germany;\\
mpt218@theorie2.physik.uni-erlangen.de}

\date{\today}

\begin{abstract}
The self-consistent separable random-phase approximation (SRPA) model with
Skyrme forces is extended to the case of magnetic excitations and applied to
the description of spin-flip and orbital M1 giant resonances in the isotopic
chain $^{142-152}$Nd. The Skyrme forces SkT6, SkM*, SLy6 and SkI3 are used. The
calculations show the onset of the scissors mode with increasing deformation. A
specific three-peak structure of the spin-flip response is found and explained
by particular neutron and proton spin-flip transitions. Although the employed
forces provide an acceptable qualitative description, the Skyrme functional
still needs further improvement to reproduce quantitatively the experimental
data.
\end{abstract}

\maketitle

\section{Introduction}

Magnetic giant resonances (GR) represent an important part of the
nuclear dynamical response \cite{Harakeh_book_01}. They were widely
investigated within various phenomenological and microscopic models,
for surveys see
\cite{Harakeh_book_01,Speth_91,Osterfeld_92,Iud_rew}. Now more
elaborated Skyrme, Gogny, and relativistic mean-field approaches
\cite{Ben,Vre05aR,Sto07aR} based on density functional theory (DFT)
are at our disposal. They are actively used for the description of
nuclear ground states and dynamics covering mainly
electrical modes but not magnetic ones. However magnetic
modes could be very useful for analyzing and improving these
approaches in the parts related to spin densities. Besides, since
these modes (especially the spin-flip M1 GR) are sensitive to the
spin-orbit splitting, they can be used to study
the spin-orbit interaction in Skyrme and Gogny forces. They can
also help to clarify the role of tensor forces and related
spin-orbit densities
\cite{Les_PRC_07_tensor,Colo_Sagawa_08,nest_PRC_09}.

This paper is devoted to an investigation of M1 spin-flip and orbital
GR within the Skyrme-Hartree-Fock (SHF) approach
\cite{Skyrme,Vau,Engel_75}.  Earlier SHF studies of these GR were
limited to a few explorations \cite{Sarriguren_M1,Hilton_98} and even
these were not fully consistent. In the study \cite{Sarriguren_M1}, a
hybrid model with partial inclusion of SHF in the Landau-Migdal
formulation was exploited while the work \cite{Hilton_98} used early
Skyrme forces and omitted the important spin density. Quite recently a
first fully self-consistent systematic SHF investigation of the
spin-flip M1 GR was performed \cite{nest_PRC_09} within the separable
Random-Phase-Approximation (SRPA) model \cite{nest_PRC_02,nest_PRC_06}
extended to magnetic excitations \cite{nest_PRC_09,Petr_PhD}. It was
shown that none of 8 different Skyrme parameterizations was able to
describe simultaneously the one-peak structure in doubly-magic nuclei
together with the two-peaks in deformed nuclei.

In this paper, the SRPA for magnetic excitations
is presented in more detail and applied to an investigation
of spin-flip M1 GR in the chain of isotopes $^{142-152}$Nd.  The
trends with increasing deformation and the number of neutrons are
explored. Besides, we present our first results for the orbital
scissors mode \cite{Iud}. The calculations employ the Skyrme
parameterizations SkT6 \cite{skt6}, SkM* \cite{skms}, SLy6
\cite{sly46}, and SkI3 \cite{ski3}.

\section{SRPA for magnetic excitations}
\subsection{General formalism}

SRPA is a fully self-consistent DFT model  where both the static mean field and
residual interaction are derived from the same functional. The present nuclear
application is based on the Skyrme functional \cite{Skyrme,Vau,Engel_75}. It
was first derived for electric excitations \cite{nest_PRC_02,nest_PRC_06} and
then extended to magnetic modes \cite{nest_PRC_09,Petr_PhD}. The
self-consistent factorization of the residual interaction in SRPA considerably
reduces the computational expense while maintaining a high accuracy. This makes
the model very suitable for systematic studies. The SRPA residual interaction
includes all contributions from the initial Skyrme functional as well as the
Coulomb (direct and exchange) and pairing (at BCS level) terms.  The model was
widely used for the investigation of electrical GR in spherical and deformed
(heavy and super-heavy) nuclei
\cite{nest_PRC_02,nest_PRC_06,nest_ijmp,nest_PRC_08}.

Starting point is the Skyrme functional with the energy density
\cite{Ben,Sto07aR}
\begin{eqnarray}
  \mathcal{H}_\mathrm{Sk}
  &=&
  \frac{b_0}{2} \rho^2- \frac{b'_0}{2} \sum_q\rho_{q}^2
  + \frac{b_3}{3} \rho^{\alpha+2}
  - \frac{b'_3}{3} \rho^{\alpha} \sum_q \rho^2_q
\nonumber
\\
 &&
 +b_1 (\rho \tau - \textbf{j}^2)
 - b'_1 \sum_q(\rho_q \tau_q - \textbf{j}^2_q)
 - \frac{b_2}{2} \rho\Delta \rho
 + \frac{b'_2}{2} \sum_q \rho_q \Delta \rho_q
\nonumber
\\
 &&
 - b_4 (\rho \nabla\textbf{J}\!+\!(\nabla\!\times\!\textbf{j})\!\!\cdot\!\!\textbf{s})
 - b'_4 \sum_q (\rho_q \nabla\textbf{J}_q\!
 +\!(\nabla\!\times\!\textbf{j}_q)\!\!\cdot\!\!\textbf{s}_q)
\nonumber
\\
  &&
  + \frac{\tilde{b}_0}{2} \textbf{s}^2
  - \frac{\tilde{b}'_0}{2} \sum_q \textbf{s}_{q}^2
+ \frac{\tilde{b}_3}{3} \rho^{\alpha} \textbf{s}^2
- \frac{\tilde{b}'_3}{3} \rho^{\alpha} \sum_q \textbf{s}^2_q
 -\frac{\tilde{b}_2}{2} \textbf{s} \!\cdot\!
  \Delta \textbf{s} + \frac{\tilde{b}'_2}{2}
  \sum_q \textbf{s}_q \!\cdot\!\Delta \textbf{s}_q
\nonumber
\\
 &&
  +\gamma_\mathrm{T}(\tilde{b}_1
   (\textbf{s}\!\cdot\!\textbf{T}\!-\!\textbf{J}^2)
  + \tilde{b}'_1
   \sum_q (\textbf{s}_q\!\cdot\!\textbf{T}_q
    \!-\!\textbf{J}_q^2))
\label{eq:skyrme_funct}
\end{eqnarray}
where $\alpha$, $\gamma_\mathrm{T}$, $b_i$, $b'_i$, $\tilde{b}_i$,
$\tilde{b}'_i$ are the force parameters.  This functional involves time-even
(nucleon $\rho_q$, kinetic-energy $\tau_q$, spin-orbit $\textbf{J}_q$) and
time-odd (current $\textbf{j}_{ q}$, spin $\textbf{s}_q$, and vector
kinetic-energy $\textbf{T}_q$) densities where $q$ denotes protons and
neutrons. Densities without index, like $\rho = \rho_p + \rho_n$, are total.
The contributions with $b_i$ (i=0,1,2,3,4) and $b'_i$ (i=0,1,2,3) are the
standard terms responsible for ground state properties and electric excitations
of even-even nuclei \cite{Ben,Sto07aR}. In the standard SHF, the isovector
spin-orbit interaction is linked to the isoscalar one by $b'_4=b_4$. The tensor
spin-orbit terms $\propto\gamma_\mathrm{T}$ are often skipped.  They can be
switched in (\ref{eq:skyrme_funct}) by the parameter $\gamma_\mathrm{T}$.  The
spin terms with $\tilde{b}_i, \tilde{b}'_i$ are relevant for odd nuclei and
magnetic modes in even-even nuclei.  Though $\tilde{b}_i, \tilde{b}'_i$ may be
uniquely determined as functions of ${b}_i,{b}'_i$ \cite{Sto07aR}, their values
were not yet well tested by nuclear data.
As was shown \cite{nest_PRC_09}, just these spin terms are of a paramount
importance for the spin-flip M1.

The general SRPA formalism is given elsewhere
\cite{nest_PRC_02,nest_PRC_06}. We present here only the basics and
peculiarities of magnetic excitations. The SRPA simplifies the
residual interaction of Skyrme RPA in a factorized (separable) form as
\begin{equation} \label{V_sep}
  \hat{V}_{\rm res}^{\rm sep}
  =
  -\frac{1}{2}\sum_{qq'}\sum_{k, k'=1}^{K}
   \{ \kappa_{qk,q'k'} {\hat X}_{qk} {\hat X}_{q'k'}
     + \eta_{qk,q'k'} {\hat Y}_{qk} {\hat Y}_{q'k'} \}
\end{equation}
where the indices $q$ and $q'$ label neutrons and protons, $k$ numbers
the separable terms, $\kappa_{qk,q'k'}$ and $\eta_{qk,q'k'}$ are the
corresponding strength matrices, and ${\hat X}_{qk}$ and ${\hat
Y}_{qk}$ are time-even and time-odd hermitian one-body operators. We
need these two kinds of the operators since the relevant Skyrme
functionals involve both time-even and time-odd densities, see
\cite{Ben,nest_PRC_02,nest_PRC_06,Petr_PhD}.

The model uses the Skyrme functional
$E[J_{q}^{\alpha}({\textbf{r}},t)]=\int d\textbf{r}
\mathcal{H}_\mathrm{Sk}(\textbf{r},t)$ depending on the  local
densities $J_{q}^{\alpha} \equiv
(\rho_q,\tau_q,\mathbf{J}_q,\mathbf{j}_q,\mathbf{s}_q,\mathbf{T}_q)$.
The separable operators and strength matrices in (\ref{V_sep}) are
self-consistently derived from this functional
and read \cite{nest_PRC_02,nest_PRC_06}
\begin{eqnarray}
\label{eq:X}
\hat{X}_{q k} &=& \sum_{q'}\hat{X}_{q k}^{q'} = i\sum_{\alpha' \alpha q'}
\frac{\delta^2 E} {\delta J_{q'}^{\alpha'} \delta J_{q}^{\alpha}} \langle
[\hat{P}_{q k} ,{\hat J}_{q}^{\alpha}] \rangle {\hat J}_{q'}^{\alpha'},
\\
\label{eq:Y} \hat{Y}_{q k} &=& \sum_{q'}\hat{Y}_{q k}^{q'} = i\sum_{\alpha'
\alpha q'} \frac{\delta^2 E} {\delta J_{q'}^{\alpha'} \delta J_{q}^{\alpha}}
\langle [\hat{Q}_{q k} ,{\hat J}_{q}^{\alpha}] \rangle {\hat J}_{q'}^{\alpha'},
\end{eqnarray}
\label{eq:kappa-eta}
\begin{eqnarray}
\label{eq:kappa}
  \kappa_{q'k',qk}^{-1 }
  &=&
  \sum_{\alpha \alpha'}
  \frac{\delta^2 E}{\delta J_{q'}^{\alpha'}\delta J_{q}^{\alpha}}
  \langle [\hat{P}_{q k},{\hat J}_{q}^{\alpha}] \rangle
  \langle [\hat{P}_{q' k'},{\hat J}_{q'}^{\alpha'}] \rangle ,
\\
\label{eq:eta}
  \eta_{q'k',qk}^{-1 }
  &=&
 \sum_{\alpha \alpha'}
  \frac{\delta^2 E} {\delta J_{q'}^{\alpha'}\delta J_{q}^{\alpha}}
  \langle [\hat{Q}_{q k},{\hat J}_{q}^{\alpha}] \rangle
  \langle [\hat{Q}_{q' k'},{\hat J}_{q'}^{\alpha'}] \rangle \: .
\end{eqnarray}
Here $\hat{J}_{q}^{\alpha}$ are the operators associated with the
local densities ${J}_{q}^{\alpha}$. Further,
$\hat{Q}_{qk}(\textbf{r})$ and $\hat{P}_{qk}(\textbf{r})$ are
generalized coordinate (time-even) and momentum (time-odd) hermitian
one-body input operators which serve as generators of the separable
terms \cite{nest_PRC_02,nest_PRC_06}.  For E$\lambda$ modes, the input
operators ${\hat Q}_{qk}(\textbf{r})$ are chosen and ${\hat
P}_{qk}(\textbf{r})=i[\hat{H},\hat{Q}_{qk}]$ where $\hat{H}$ stands
for the full Hamiltonian. Instead, for M$\lambda$ modes, the input
operators ${\hat P}_{qk}(\textbf{r})$ are chosen and ${\hat
Q}_{qk}(\textbf{r})=i[\hat{H},\hat{P}_{qk}]$ are deduced.

The RPA equations for coordinate $\bar{q}_{qk}$ and momentum
$\bar{p}_{qk}$ variables read
\begin{eqnarray}
\label{eq:RPA_1} \sum_{qk} \{ \bar{q}_{qk}^{\nu} [F_{q'k',qk}^{(XX)}-
\kappa_{q'k',qk}^{-1}] +\bar{p}_{qk}^{\nu} F_{q'k',qk}^{(XY)} \} &=& 0 ,
\\
\label{eq:RPA_2} \sum_{qk} \{ \bar{q}_{qk}^{\nu} F_{q'k',sk}^{(YX)}
+\bar{p}_{qk}^{\nu} [F_{q'k',qk}^{(YY)} - \eta_{q'k',qk}^{-1}] \} &=& 0
\end{eqnarray}
with
\begin{equation}
\label{eq:F_AB}
  F_{q'k',qk}^{(AB)}
 = 2 \sum_{q", ph \in q"}\alpha_{AB}
\frac{\langle ph|\hat{A}^{q"}_{qk} \rangle^*
   \langle ph|\hat{B}^{q"}_{q'k'} \rangle}
   {\varepsilon_{ph}^2-\omega_{\nu}^2}
\end{equation}
and
\begin{equation}
\alpha_{AB}=\left(
\begin{array}{l}
{\varepsilon_{ph}, \quad\mbox{for} \; \hat{A}=\hat{B}}\\
{-i\omega_{\nu}, \; \mbox{for} \; \hat{A}=\hat{Y}, \hat{B}=\hat{X}}\\
{i\omega_{\nu}, \quad \mbox{for} \; \hat{A}=\hat{X}, \hat{B}=\hat{Y}}
\end{array}
\right) \; .
\end{equation}
Here $\langle ph|\hat{A}^{q"}_{q'k'} \rangle$ is the matrix element for the
two-quasiparticle state $|ph\rangle$, $\varepsilon_{ph}$ is the energy of this
state, $\omega_{\nu}$ is the energy of the RPA state $|\nu\rangle$. The RPA
phonon operator reads
\begin{equation}
  \hat{C}^{\dagger}_{\nu}
  =
  \sum_q \sum_{ph\in q}
  \left(c^{\nu -}_{ph}\hat{d}^{\dagger}_{ph}
   - {c^{\nu +}_{ph}}_{\mbox{}}\hat{d}^{\mbox{}}_{ph}\right)
\label{eq:geneigen}
\end{equation}
where $\hat{d}^\dagger_{ph}$ ($\hat{d}_{ph}$) are the creation
(destruction) operators of two-quasiparticle configurations with the
amplitudes determined by solutions of
(\ref{eq:RPA_1})-(\ref{eq:RPA_2}):
\begin{eqnarray}
  c^{\nu \pm}_{ph \in q}
  =
  -\sum_{q'k'}
  \frac{\bar{q}_{q'k'}^{\nu}
        \langle ph|\hat{X}^q_{q'k'}\rangle
        \mp i
        \bar{p}_{q'k'}^{\nu}  \langle ph|\hat{Y}^q_{q'k'}\rangle}
  {2(\varepsilon_{ph}\pm\omega_{\nu})} \; .
\label{eq:c_pm_qp}
\end{eqnarray}

Following (\ref{eq:X})-(\ref{eq:eta}), the separable ansatz
(\ref{V_sep}) explores the residual interaction of the Skyrme
functional through the second functional derivatives. The calculations
show that, for spin-flip magnetic modes, the spin
\begin{eqnarray}
\frac{\delta^2 E}{\delta \textbf{s}_{q'}(\textbf{r}') \delta
\textbf{s}_{q}(\textbf{r})} &=& \Biggl[ \tilde{b}_0 - \tilde{b}'_0 \delta_{q
q'} + \tilde{b}_3 \frac{2}{3} \rho^{\alpha}(\textbf{r})
- \frac{2}{3} \tilde{b}'_3 \rho^{\alpha}(\textbf{r}) \delta_{q q'}
\\
&-&
(\tilde{b}_2 - \tilde{b}'_2 \delta_{q q'}) \Delta_{\textbf{r}} \Biggr]
\delta(\textbf{r}-\textbf{r}') \; , \nonumber
\end{eqnarray}
spin-orbit
\begin{eqnarray}
\frac{\delta^2 E}{\delta \textbf{J}_{q'}(\textbf{r}') \delta
\rho_{q}(\textbf{r})} &=& (b_4 + b'_4 \delta_{q q'}) \nabla_{\textbf{r}}
\delta(\textbf{r}-\textbf{r}')
\; ,
\\
\frac{\delta^2 E}{\delta \textbf{j}_{k;q'}(\textbf{r}') \delta
\textbf{s}_{l;q}(\textbf{r})} &=& (b_4 + b'_4 \delta_{q q'}) (\varepsilon_{klm}
\nabla_{m;\textbf{r}}) \delta(\textbf{r}-\textbf{r}') \; , \nonumber
\end{eqnarray}
and tensor terms
\begin{eqnarray}
\frac{\delta^2 E}{\delta \textbf{J}_{q'}(\textbf{r}') \delta
\textbf{J}_{q}(\textbf{r})} &=& - 2(\tilde{b}_1 + \tilde{b}'_1 \delta_{q q'})
\delta(\textbf{r}-\textbf{r}') \; ,
\\
\frac{\delta^2 E}{\delta \textbf{T}_{q'}(\textbf{r}') \delta
\textbf{s}_{q}(\textbf{r})} &=& (\tilde{b}_1 + \tilde{b}'_1 \delta_{q q'})
\delta(\textbf{r}-\textbf{r}')
\end{eqnarray}
are most important. In deformed nuclei, the magnetic modes
couple to electric motion \cite{ML_EL}. Then the relevant electric terms
\cite{nest_PRC_02,nest_PRC_06} should be added.
For the scissors mode, which exists only in deformed nuclei,
the spin-orbit, tensor, and electric terms are essential.

As was mentioned above, the model is self-consistent in the sense that both the
static mean field
\begin{eqnarray}\label{eq:h_0}
  \hat{h}_0 =
  \sum_{\alpha q}
  \frac{\delta E} {\delta J_{q}^{\alpha}}\hat{J}_{q}^{\alpha}
\end{eqnarray}
and the residual interaction (\ref{V_sep}) are derived from the same
functional. The rank of the RPA matrix (\ref{eq:RPA_1})-(\ref{eq:RPA_2}) is
determined by the number $K$ of the input $\hat{Q_{qk}}$ or $\hat{P_{qk}}$
operators. Usually $K=2 \div 5$ and so the rank is small
\cite{nest_PRC_06,nest_ijmp,nest_PRC_08}.  This reduces the computational
effort and allows systematic studies even for heavy deformed nuclei.

The pairing functional reads $V_\mathrm{pair}=1/2\sum_q
G_q\chi_q^{\mbox{}}\chi^*_q$ where $\chi_q$ is the pairing density and $G_q$ is
the pairing strength \cite{Sto07aR}. In the present study, the pairing is
included at the BCS level through the quasiparticle energies and Bogoliubov
coefficients.

\subsection{Strength function}

GR in heavy nuclei are formed by many RPA states whose detailed
structures cannot be resolved experimentally. Then a direct
computation of the strength function is more efficient and
reasonable. In SRPA the strength function for M1 excitations reads
\begin{eqnarray}
\label{eq:strength_func}
  S(M1 ; \omega) &=& \sum_{\nu \ne 0}
  |\langle\Psi_\nu|\hat{M}\rangle|^2
  \zeta(\omega - \omega_{\nu})
\\
\nonumber &=&
   \Im\left[
   \frac{\sum_{\beta \beta'}
            F_{\beta \beta'}(z) D_{\beta}(z) D_{\beta'}(z)}
        {\pi F(z)}
  \right]_{z=\omega\!+\!i\frac{\Delta}{2}}
 + \sum_{q, ph \in q} |\langle ph|\hat{M}\rangle|^2
\zeta(\omega -\varepsilon_{ph})
\end{eqnarray}
where $\langle\Psi_\nu|\hat{M}\rangle$ is the matrix element of M1 transition
between the ground  and excited $|\Psi_\nu\rangle$ RPA states, and $
\zeta(\omega - \omega_{\nu}) =
  \Delta /[2\pi[(\omega- \omega_{\nu})^2+\Delta^2/4]]$
is the Lorentz weight  to
simulate the broadening effects beyond SRPA (escape widths,
coupling with complex configurations). In the present study,
the Lorentz averaging parameter is $\Delta$=1 MeV. Further, $\beta=qk\tau$
with $\tau$ labeling $X$ and $Y$-operators, $\Im$ means the
imaginary part of the value inside the brackets,  $F(z)$ is
the determinant of the RPA matrix (\ref{eq:RPA_1})-(\ref{eq:RPA_2})
 with $\omega_{\nu}$ replaced by the complex
argument $z$, $F_{\beta \beta'}(z)$ is the algebraic supplement of the
determinant, and
$$
\label{eq:A_X_1}
D_{qk}^{(X)}(z) = \sum_{q', ph \in {q'}}
\frac{\omega_{\nu}
      \langle ph|X^{q'}_{qk} \rangle \langle ph|\hat{M} \rangle}
      {\varepsilon_{ph}^2-z^2} ,
\;
D_{qk}^{(Y)}(z) = \sum_{q', ph \in q'}
\frac{i \varepsilon_{ph}
       \langle ph|Y^{q'}_{qk} \rangle \langle ph|\hat{M} \rangle}
      {\varepsilon_{ph}^2-z^2} .
$$
The operator of M1 transition reads
${\hat M}(M1\mu) =
\mu_B \sqrt{3/(8\pi)}\sum_q [g^{q}_s {\hat s}_{\mu} + g^{q}_l {\hat l}_{\mu}] $
where spin $g$-factors $g^{p}_s = 5.58 \varsigma$ and
$g^{n}_s = - 3.82 \varsigma$ are quenched by $\varsigma$=0.7 and
orbital $g$-factors are $g^{p}_l = 1$ and $g^{n}_l = 0$; ${\hat s}_{\mu}$ and
${\hat l}_{\mu}$ are spin and orbital operators.

In the present study, the strength function (\ref{eq:strength_func})
is calculated only for the branch $\mu=1$ which gives the
$K^{\pi}=1^+$ states relevant for both spin-flip and orbital modes.

\subsection{Choice of input operators}

The SRPA formalism requires to choose for M1 modes the generating
operators ${\hat P}_{qk}({\textbf r})$. Their choice is crucial for a
fast convergence of the separable expansion $\hat{V}_{\rm res}^{\rm
sep}$ to the true Skyrme residual interaction. We achieve this by
using such ${\hat P}_{qk}({\textbf r})$ which allow the separable
operators $\hat{X}_{q k}({\textbf r})$ and $\hat{Y}_{q k}({\textbf
r})$ to have maxima in different spatial regions of the nucleus, in
the surface and interior. Such a flexible choice allows to achieve
good convergence already with a few separable terms.

The physical arguments suggest that, for spin-flip mode, the leading
input operator ${\hat P}_{q1}(\textbf{r})$ should have the form of the
spin part of the applied external field (\ref{eq:M1}) with $\mu=1$,
i.e. $\hat{P}_{q1}=\hat{s}^q_+$. The detailed distributions depends on
the interplay of surface and volume excitations.  This can be resolved
by taking into account the interior of the nucleus
\cite{nest_PRC_02,nest_PRC_06,Petr_PhD} by adding input operators with
different radial parts, $\hat{P}_{q2}=r^2\hat{s}^q_+$,
$\hat{P}_{q3}=r^4\hat{s}^q_+$. This results in $\hat{X}_{q k}({\textbf
r})$ and $\hat{Y}_{q k}({\textbf r})$ interaction operators having more
sensitivity in the interior.

In deformed nuclei we should take into account the coupling between
magnetic and electric $K^{\pi}=1^+$ states. So the quadrupole input
operator ${\hat Q}_{q}=r^2 Y_{21}$ with the counterpart ${\hat
P}_{q}=i[\hat{H},\hat{Q}_{q}]$ should be added. This operator
generates quadrupole excitations in both spherical and deformed
nuclei. Besides it allows to explore the scissors mode in deformed
nuclei.

In Fig. 1, the spin-flip ($g^{n,p}_l = 0$) strength function
(\ref{eq:strength_func}) in $^{150}$Nd is plotted for different sets
of input operators. It is seen that all the sets give very similar
results. In principle, it would suffice to use the minimal set
$\hat{P}_{q}=\hat{s}^q_+$ and ${\hat Q}_{q}=r^2 Y_{21}$. However, to
be on the safe side, we will use in the following calculations a set
from 3 input operators, by adding also the operator
$\hat{P}_{q}=r^2\hat{s}^q_+$.  Then K=3 and we have the RPA matrix of
the modest rank 4K=12.
\begin{figure}[th]
\label{fig:input}
\centerline{\psfig{file=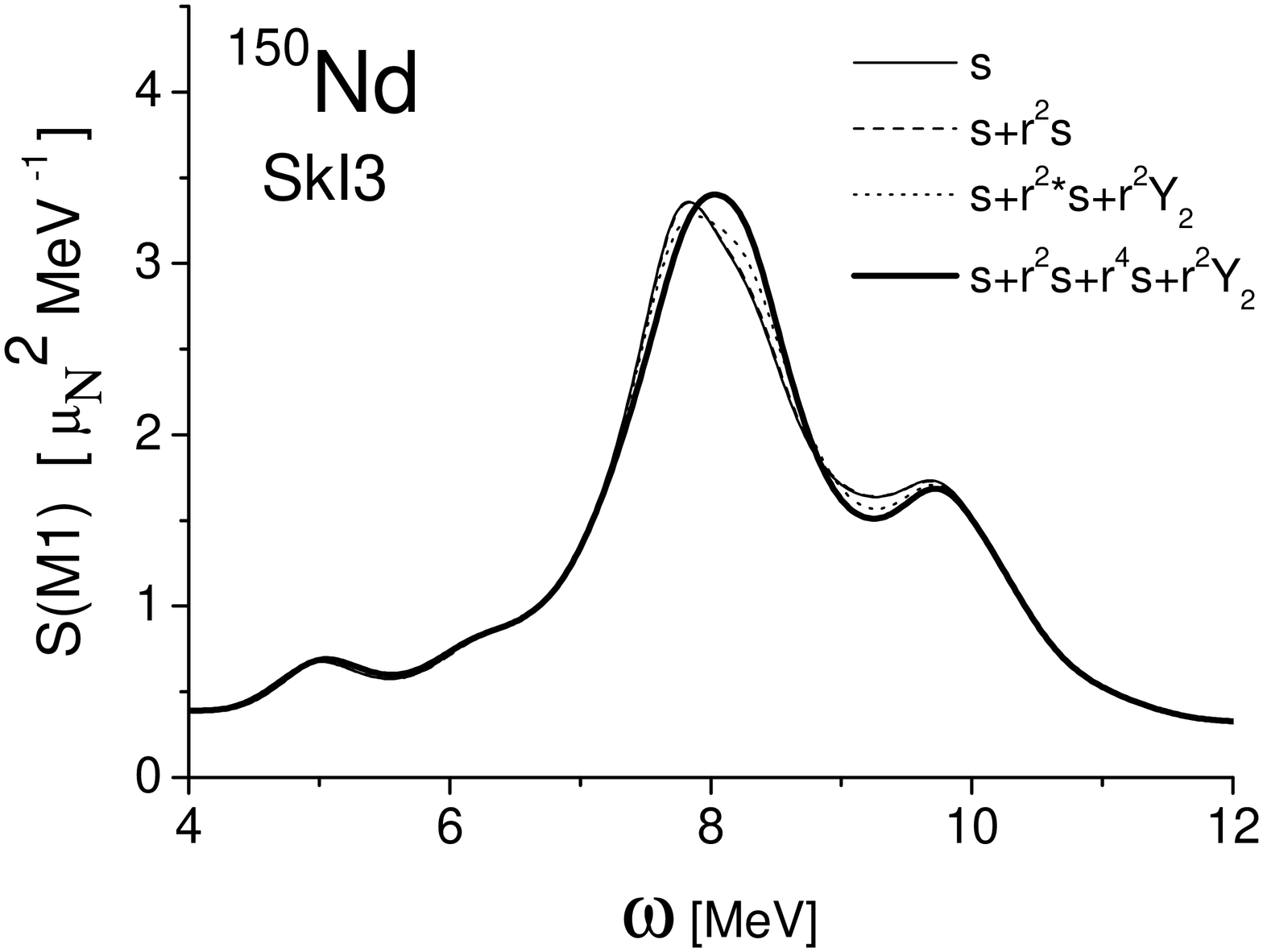,width=5.0cm,angle=-90}}
\caption{
Strength function for spin-flip M1 GR in $^{150}$Nd, calculated with
the force SkI3 for four sets of input operators, as indicated in the figure.}
\end{figure}

Note, that in terms of two-quasiparticle matrix elements, the
relations between input operators and their counterparts have the form
\begin{equation}
\nonumber \hat{P}_{qk} \to {\hat Q}_{qk}(\textbf{r})=i[\hat{H},\hat{P}_{qk}]
\to \langle ph|\hat{Q}_{qk}\rangle = 2\varepsilon_{ph} \langle
ph|\hat{P}_{qk}\rangle - \langle ph|\hat{X}_{qk}^q\rangle
\end{equation}
for magnetic modes and
\begin{equation}
\nonumber  {\hat Q}_{qk} \to {\hat P}_{qk}(\textbf{r})=i[\hat{H},\hat{Q}_{qk}]
\to \langle ph|\hat{P}_{qk}\rangle = 2\varepsilon_{ph} \langle
ph|\hat{Q}_{qk}\rangle - \langle ph|\hat{Y}_{qk}^q\rangle
\end{equation}
for electric modes. For more details see
\cite{nest_PRC_02,nest_PRC_06,Petr_PhD}.

\section{Results and discussion}

Results of the calculations are presented in Figs. 2-4.  Fig. 2
shows the spin-flip and orbital M1 strengths for the isotopes
$^{142-152}$Nd. The strengths are computed with the force SkI3 by
using $g_l^p$=0 and $g_s^{n,p}$=0, respectively.  The isotopic chain
ranges from semi-magic spherical (A=142) to axially deformed (A=150, 152)
nuclei. The soft nuclei (A=144,146,148)
require probably a beyond-RPA treatment. Nevertheless, they are
included to illustrate the trends.
\begin{figure}[th] \label{142-152Nd}
\centerline{\psfig{file=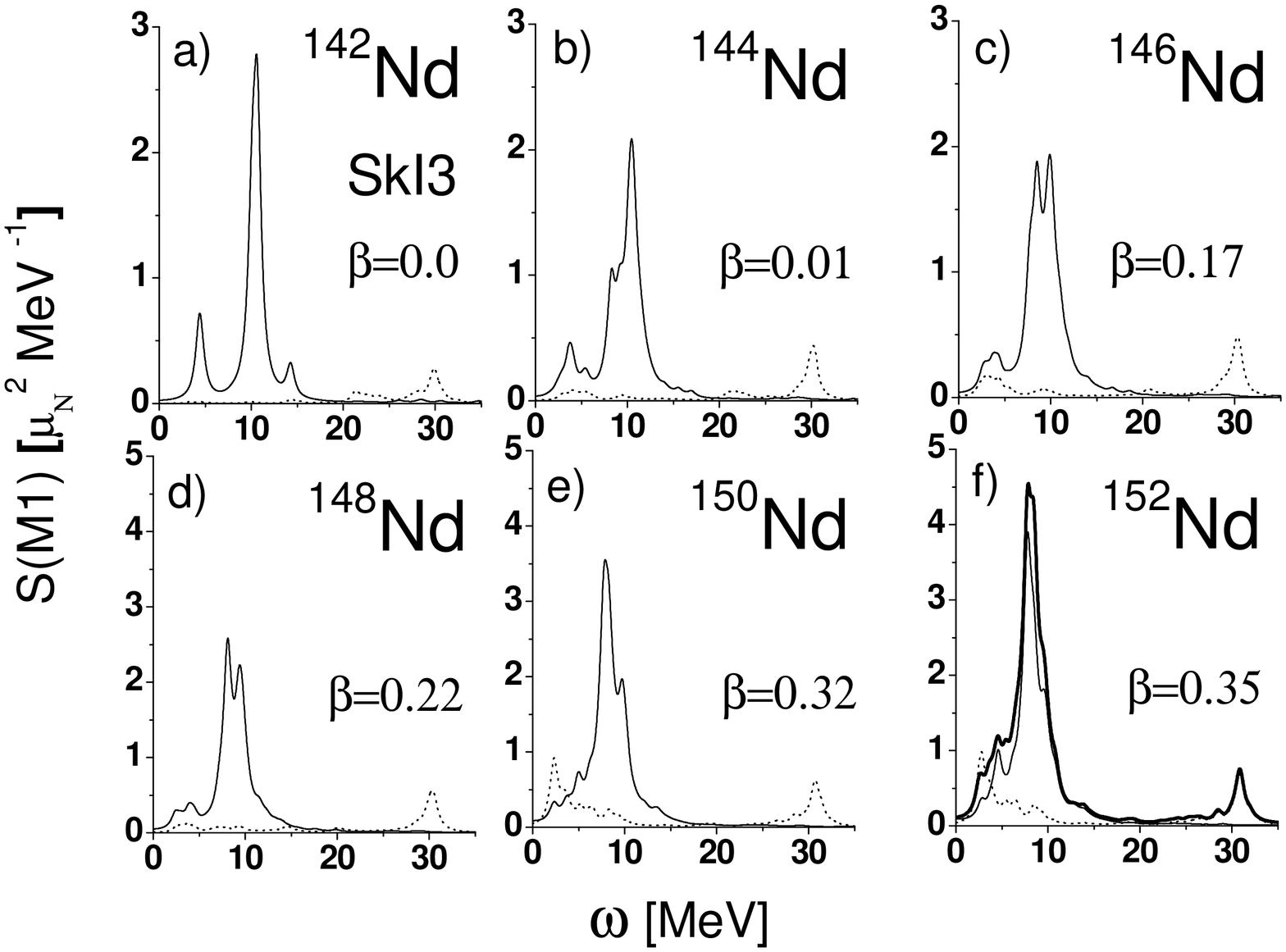,width=6.5cm,angle=-90}}
\caption{
Spin-flip (solid curve) and orbital (dotted curve) strength functions
in $^{142-152}$Nd, calculated with the force SkI3. For every isotope
the parameter of quadrupole deformation $\beta$ defined in Ref.
\protect\cite{nest_PRC_08} is shown. For $^{152}$Nd,
the total strength (bold curve) with both spin-flip and orbital
contributions is presented.}
\end{figure}
\begin{figure}[th] \label{np}
\centerline{\psfig{file=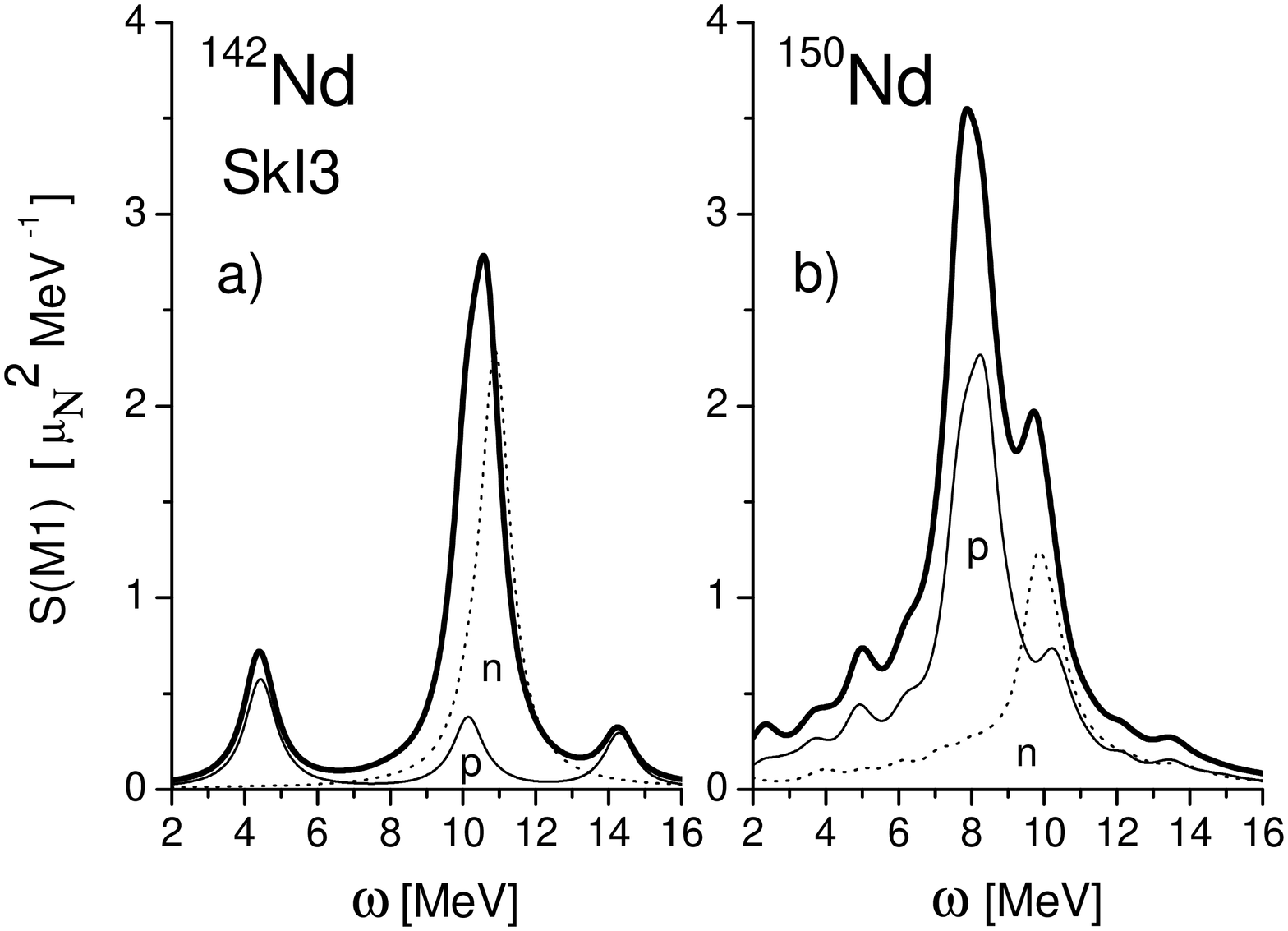,height=8.5cm,width=4.7cm,angle=-90}} \caption{
Total spin-flip (bold curve), proton (solid curve), and neutron (dash curve)
strength functions in $^{142}$Nd and $^{150}$Nd, calculated with the force
SkI3. The proton and neutron strengths are computed with $g^{n}_{s,\;l} = 0$
and $g^{p}_{s,\;l} = 0$, respectively.}
\end{figure}

Figure 2 includes the high-energy orbital strength at 30 MeV for all
the isotopes.  Actually this is the $\lambda\mu=21$ branch of the
isovector electric quadrupole GR, though with an overestimated
energy. The low-energy scissor orbital mode is absent in spherical
$^{142}$Nd but then appears and steadily increases with growth of the
deformation in heavier isotopes. In deformed $^{150,152}$Nd, this mode
dominates at 2-4 MeV and gives a significant contribution to M1
strength at 2-7 MeV. The comparison of the total, spin-flip and
orbital strengths in $^{152}$Nd shows that the interference between
spin-flip and orbital modes can be both constructive and
destructive. The results for other Skyrme forces used in the paper,
SkT6, SkM*, and SLy6, are similar.

Figure 2 also demonstrates a significant change of the structure of
the spin-flip resonance with increasing deformation and neutron
number. The distinct two-peak structure at 4 and 10 MeV in $^{142}$Nd
evolves to a broad one-peak resonance in $^{150,152}$Nd.  Following our
analysis of the structure of the RPA states, the spin-flip strength in
Nd isotopes is determined by neutron $\nu(1h_{11/2}^{-1},1h_{9/2})$
and proton $\pi(2d_{5/2}^{-1},2d_{3/2})$ as well as
$\pi(1g_{9/2}^{-1},1g_{7/2})$ transitions.  This is confirmed by
Fig. 3 where proton and neutron contributions are separated. One sees
that in $^{142}$Nd the neutron strength is given by one peak while the proton
distribution exhibits two peaks, of $\pi(2d_{5/2}^{-1},2d_{3/2})$ and
$\pi(1g_{9/2}^{-1},1g_{7/2})$ origin for 4 MeV and 10 MeV structures,
respectively. The neutron contribution to the 10 MeV peak
weakens from $^{142}$Nd to $^{150}$Nd since in the latter the
neutron subshell $1h_{9/2}$ is already partly occupied. Instead, the proton
contribution is concentrated and enforced.

Note that in Fig. 3 the proton and neutron peaks at 10 MeV are
close to each other and actually form one broad resonance. This is a
consequence of close neutron and proton spin-orbit splittings
provided by SkI3 for medium and heavy nuclei \cite{ski3}. For the comparison,
the forces SkM* and SLy6, with more different neutron and proton spin-orbit
splittings \cite{nest_PRC_09,skms,sly46} give a stronger splitting of the
resonance in both $^{142}$Nd and $^{150}$Nd (Fig. 4b,d).  Thus the M1
strength distribution is related to the spin-orbit structure of
the underlying single-particle states.

The above consideration indicates that the structure of spin-flip M1
mode depends on the Skyrme force. This is further illustrated in
Fig. 4 where the results for four Skyrme forces are presented. The
forces yield quite different results. Besides, none of these
forces provides a proper quantitative description of the experimental
data for $^{150}$Nd \cite{exp_Nd}.  In particular, the high spike at 6
MeV is not reproduced. This confirms the finding \cite{nest_PRC_09}
that most of Skyrme parameterizations do not fully describe the
spin-flip spectra. Nevertheless, our calculations allow to suggest a
possible origin of three M1 peaks observed in nuclides at the onset
of rare-earth region \cite{exp_Nd,exp_Gd}. Following the above
analysis, they can be considered as $\pi(2d_{5/2}^{-1},2d_{3/2})$,
$\pi(1g_{9/2}^{-1},1g_{7/2})$, and $\nu(1h_{11/2}^{-1},1h_{9/2})$
spin-flip transitions.
\begin{figure}[th] \label{forces}
\centerline{\psfig{file=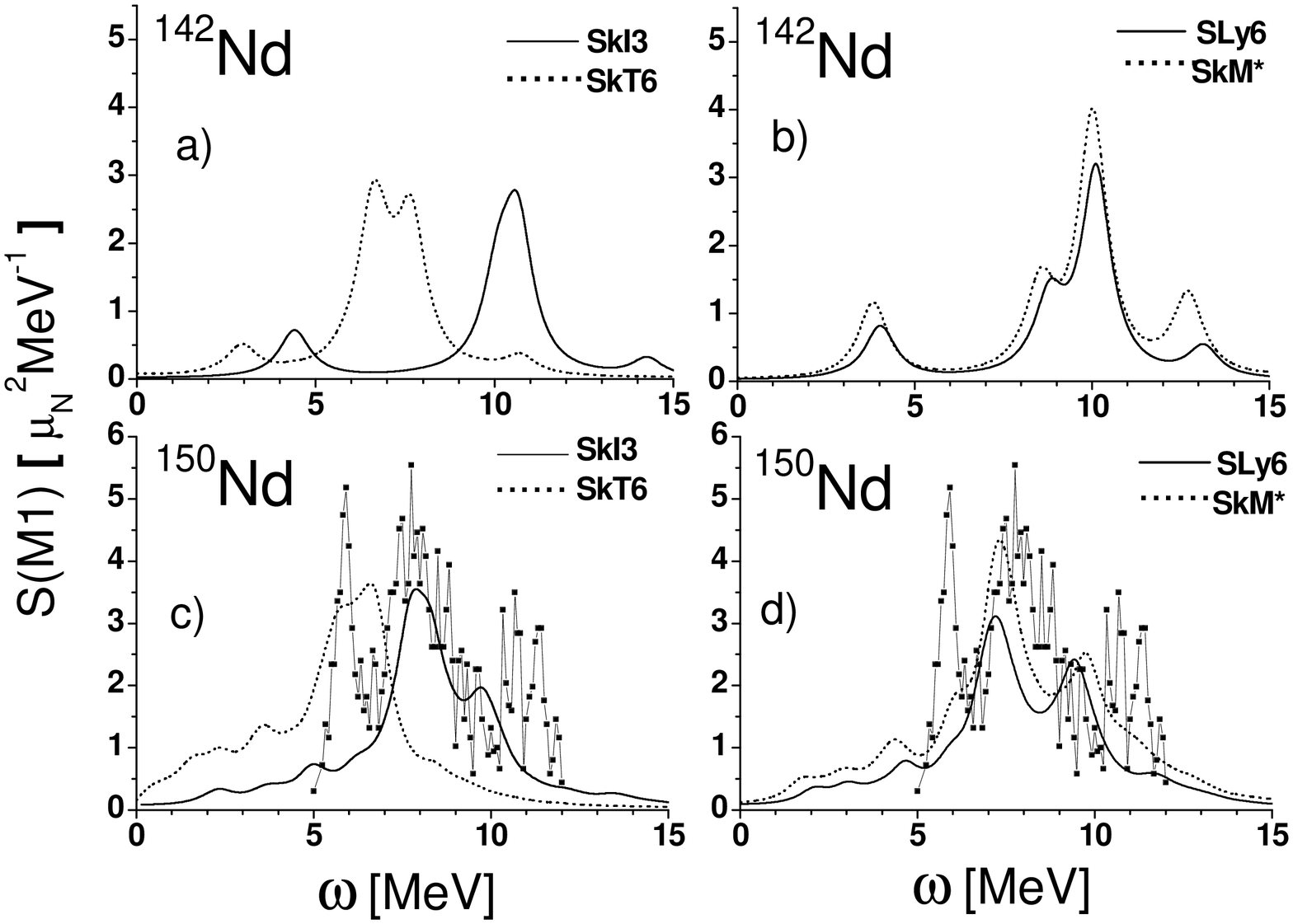,height=9.2cm,width=6.8cm,angle=-90}} \caption{
Total strength function in $^{142}$Nd and $^{150}$Nd for the Skyrme forces
SkT6,SkI3, SkM*, and SLy6.
In $^{150}$Nd the experimental data \protect\cite{exp_Nd} are depicted.}
\end{figure}

\section{Conclusions}

The orbital and spin-flip M1 giant resonances in Nd isotopes
with A=142, 144, 146, 148, 150, and 152 were investigated within the
self-consistent separable random-phase-approximation model (SRPA)
\cite{nest_PRC_02,nest_PRC_06}.
The calculations show the appearance of the scissors mode while moving
from the spherical $^{142}$Nd to deformed $^{152}$Nd.  The evolution
of the spin-flip strength with deformation and neutron number was
analyzed. The observed three-bump structure of this strength is
explained by contributions of $\pi(2d_{5/2}^{-1},2d_{3/2})$,
$\pi(1g_{9/2}^{-1},1g_{7/2})$, and $\nu(1h_{11/2}^{-1},1h_{9/2})$
spin-flip transitions. It is shown that the largest M1 peak in
$^{142}$Nd is mainly determined by the neutron contribution.
The calculations reveal a still poor
quantitative description of spin-flip M1 resonance by present-days
Skyrme parameterizations. Further improvement of Skyrme forces is
necessary, especially of their spin-orbit parts.

\section*{Acknowledgments}
The work was supported by the DFG RE-322/12-1, Heisenberg-Landau
(Germany - BLTP JINR), and Votruba - Blokhintsev (Czech Republic - BLTP JINR)
grants. W.K. and P.-G.R. are grateful for the BMBF support under contracts 06
DD 9052D and 06 ER 9063. Being a part of the research plan MSM 0021620859
(Education Ministry of Czech Republic) this work was also funded by
Czech grant agency (grant No. 202/06/0363). P.V. is grateful for the FIDIPRO
support.

\end{document}